\documentclass[aps,prb,amsfonts,amssymb,amsmath,twocolumn,floatfix]{revtex4}

\usepackage{graphicx}
\usepackage{amsmath}

\begin{document}

\title{Impurity Scattering and Mott's Formula in Graphene}

\author{Tomas L\"ofwander and Mikael Fogelstr\"om}

\affiliation{Department of Microtechnology and Nanoscience,
Chalmers University of Technology, S-412 96 G\"oteborg, Sweden}

\date{\today}

\begin{abstract}

  We present calculations of the thermal and electric linear response
  in graphene, including disorder in the self-consistent t-matrix
  approximation. For strong impurity scattering, near the unitary
  limit, the formation of a band of impurity states near the Fermi
  level leads to that Mott's relation holds at low temperature. For
  higher temperatures, there are strong deviations due to the linear
  density of states. The low-temperature thermopower is proportional
  to the inverse of the impurity potential and the inverse of the
  impurity density. Information about impurity scattering in graphene
  can be extracted from the thermopower, either measured directly, or
  extracted via Mott's relation from the electron-density dependence
  of the electric conductivity.

\end{abstract}

\maketitle


The recent isolation of the two-dimensional one-atom thick honeycomb
crystal of carbon,\cite{nov04} called graphene, has generated
tremendous interest. Partly because of the potential of carbon-based
nano-scale electronics, but also for fundamental
reasons.\cite{nov05b,zha05,geim07R,ber06,fer06,hee07,bos07,sch07,chen07,tra07,ryc07,che07,son06,sil07,kat07R}
Recent experiments\cite{nov05b,zha05,geim07R} have shown that the
charge conductance of graphene reaches a universal value $4e^2/h$ at
low temperatures. It has also been found that the conductivity is
linearly dependent on the electron density. Furthermore, an
unconventional half-integer quantum Hall effect has been discovered.

All these results are nicely in line with theoretical results based on
an effective low-energy theory of graphene within which charge
carriers are massless Dirac
fermions.\cite{wal47,slo58,sho98,gus05b,gus06,per06,ost06,zie06,nom07,das07,ando05R,kat07R}
The low-temperature conductivity of graphene is most probably limited
by scattering against impurities, such as vacancies or other
imperfections.\cite{per06,gus06,ost06,zie06,nom07} For strong
scattering, in the unitary limit, a band of impurity states centered
at the Dirac point is formed.\cite{mah04,pog06,per06,ost06,weh07} For
small temperatures, compared with the impurity band width, the
conductivity is predicted\cite{sho98,gus05b,gus06,per06,ost06} to
reach a minimal universal value $4e^2/(\pi h)$. Within the same model
of strong impurity scattering, the conductivity varies linearly with
charge density for chemical potential shifts that are large compared
with the impurity band width.\cite{ost06} In contrast, linear
dependence is absent for weak impurity scattering.\cite{ost06} In a
slightly different scenario, Coulomb randomly distributed scatterers
are predicted to lead to similar properties.\cite{nom07}

In this paper we explore further the consequences of impurity
scattering near the unitary limit, and present results for the
response to a thermal gradient in addition to the more widely studied
electric response. In particular, we find that the thermopower can
provide valuable information about impurities in graphene.

The related Nernst signal, that appears in a magnetic field, was
studied theoretically in Ref.~\onlinecite{gus06}. Further properties
of the thermopower in a magnetic field was studied in
Ref.~\onlinecite{dor07} in the strict unitary limit for which the
thermopower vanishes at charge neutrality. If vacancies are strong
scattering centers, the impurity potential is large, estimated e.g. by
graphene's vacuum level. Near unitary scattering can occur, although
the strict unitary limit can in reality not be reached. This has a big
impact on the thermopower, as the impurity band is shifted from the
Dirac point which leads to a large electron hole
asymmetry.\cite{mah04,pog06,weh07} Here we explore the
consequences of such asymmetry in detail with the goal of extracting
information about impurities in graphene from transport.


The starting point for the calculations is the tight-binding model for
clean graphene\cite{wal47,slo58}
\begin{equation}
H_0 = -t \sum_{<ij>}\left( a_{i}^{\dagger} b_j + b_{j}^{\dagger} a_i \right),
\end{equation}
where $t$ is the nearest neighbor hopping amplitude. The operators
$a_i^{\dagger}$ and $b_j^{\dagger}$ creat electrons on sites $i$ and
$j$ in the graphene honeycomb lattice. The fact that this lattice can
be considered as two displaced triangular lattices, with a unit cell
consisting of two atoms, here denoted $A$ and $B$, is made explicit by
the introduction of the two creation operators. In reciprocal space,
the Fermi surface is reduced to two inequivalent so-called K-points at
the corners of the first Brillouin zone, which we find at ${\bf
  K}_{\nu}=4\pi\nu {\hat k}_x/(3\sqrt{3}a)$, where $a$ is the nearest
neighbor distance, $\nu=\pm 1$, and ${\hat k}_x$ is unit vector along
the $k_x$-axis. At low energies $\epsilon\ll t$, the dispersion is
linear around these points, $\epsilon({\pmb\kappa})\approx\pm
v_f|{\pmb\kappa}|$, where ${\pmb\kappa}$ is the k-vector measured
relative to the K-point, and $v_f=3at/2$ is the Fermi velocity. The
clean limit retarded Green's function in the K-point (low-energy)
approximation is the inverse of the $2\times 2$ (in the space of the
atoms $A$ and $B$) Dirac Hamiltonian matrix and has the
form\cite{per06}
\begin{equation}
\hat G_{\nu}^{R(0)}({\pmb\kappa},\epsilon) = \frac{1}{2} \sum_{\lambda=\pm 1}
\frac{1}{\epsilon-\lambda v_f|{\pmb\kappa}|+i0^+}
\left(
\begin{array}{cc}
1 & \lambda\nu e^{-i\nu\beta}\\
\lambda\nu e^{i\nu\beta} & 1
\end{array}
\right),\nonumber
\end{equation}
where $\beta=\mbox{arg}(\kappa_x+i\kappa_y)$ is an angle defining the
direction of the vector ${\pmb\kappa}$ with respect to the
$k_x$-axis.

We include impurities by adding
\begin{equation}
H_{\sf imp} = \sum_{i=1}^{N_{\sf imp}^A} V_{\sf imp} a_i^{\dagger}a_i
        +\sum_{j=1}^{N_{\sf imp}^B} V_{\sf imp} b_j^{\dagger}b_j,
\end{equation}
to the Hamiltonian, where $V_{\sf imp}$ is the impurity strength. The
number of impurities $N_{\sf imp}$ in the two sub-lattices A and B is
assumed approximately equal and small compared with the number $N$ of
unit cells in the crystal. In the dilute limit, when we only keep
terms of first order in the density $n_{\sf imp}=N_{\sf imp}/N$, crossing
diagrams are neglected when the configuration average is performed
over the random distribution of impurities.\cite{MahanBook} The
resulting self-energy is written in terms of a t-matrix
$\hat\Sigma^R(\epsilon)=n_{\sf imp}\hat T^R(\epsilon)$, where
\begin{equation}
\hat T^R(\epsilon) = V_{\sf imp}\left[ {\hat 1}
- (V_{\sf imp}/N)\sum_{\bf k} \hat G^R(\bf k,\epsilon) \right]^{-1}.
\end{equation}
The sum over ${\bf k}$ can be performed analytically in the K-point
approximation. The off-diagonal components vanish, the resulting
self-energy is diagonal, and the average Green's function is
%
$
(1/N)\sum_{\bf k} \hat G^R({\bf k},\epsilon) = \hat 1 (z/\epsilon_c^2)
\log\left[-z^2/(\epsilon_c^2-z^2)\right],
$
%
where $z=\epsilon-\Sigma^R(\epsilon)$. The energy cut-off $\epsilon_c$
is related to the cut-off $k_c$ in reciprocal space below which the
dispersion is linear.

\begin{figure}[t]
\includegraphics[width=\columnwidth]{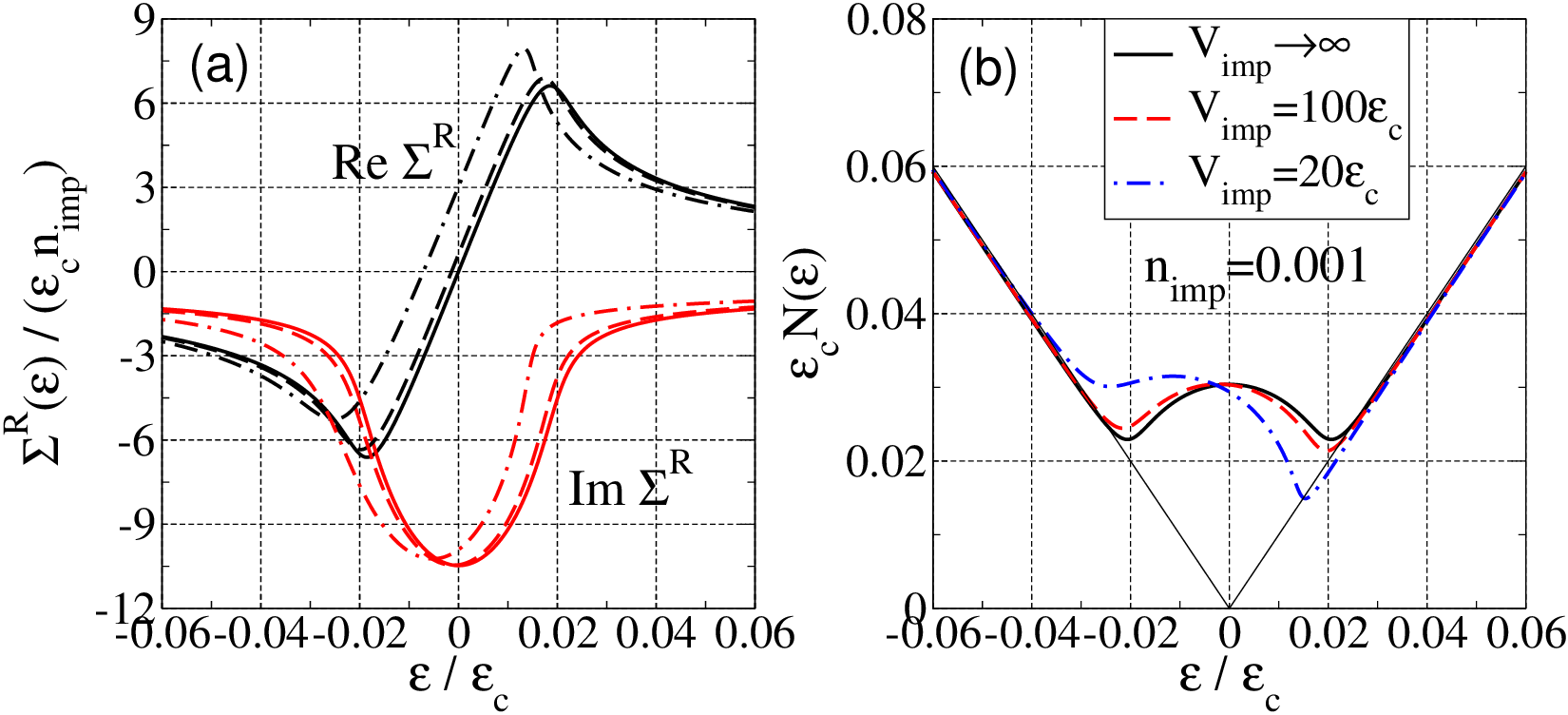}
\caption{(a) The impurity self energy and (b) the density of states
  for large impurity potential $V_{\sf imp}$. An impurity band of
  width $\gamma\sim\epsilon_c\sqrt{n_{\sf imp}}$ is formed near the
  Fermi level at an energy $\epsilon_r\sim\epsilon_c^2/V_{\sf
    imp}$. For $\epsilon\gg\gamma$ the spectrum is linear.}
\label{fig:dos}
\end{figure}

We plot the self-consistent self-energy and the resulting density of
states in Fig.~\ref{fig:dos}. In the unitary limit, defined as
$V_{\sf imp}\rightarrow\infty$, a band of impurity states is formed
centered at the Dirac point.\cite{mah04,pog06,per06,ost06,weh07} The
band width, estimated by $\gamma=-\Im\Sigma^R(0)$, is computed by
solving the equation
$(2\gamma^2/\epsilon_c^2)\ln(\epsilon_c/\gamma)=n_{\sf imp}$. To
logarithmic accuracy, $\gamma$ scales as $\sqrt{n_{\sf imp}}$. For
$V_{\sf imp}$ deviating from the unitary limit, the impurity band is
shifted away from the Fermi level to an energy $\epsilon_r$. The
self-energy is approximated as
%
$
\Sigma^R(\epsilon) = -i\gamma + \alpha(\epsilon-\epsilon_r) + ...
$
%
for small energies $\epsilon-\epsilon_r\ll\gamma$. To lowest order, we
have $\epsilon_r=-\epsilon_c^2/(2V_{\sf imp}[\ln(\epsilon_c/\gamma)-1])$
and $\alpha=\alpha_0/(1+\alpha_0)$, where
$\alpha_0=(1/n_{\sf imp})(2\gamma^2/\epsilon_c^2)[\ln(\epsilon_c/\gamma)-1]$
is only weakly (logarithmically) dependent on the density of
impurities. The shift of the impurity band away from the Fermi level
by the amount $\epsilon_r$ leads to a large electron-hole asymmetry
and the anomalous thermoelectric response that we study below. For
large energies, $\epsilon\gg\gamma$, the density of states is
essentially the same as in the clean limit.

Nearest neighbor hopping ($t^{\prime}$) give an electron-hole
asymmetric contribution to the dispersion, but it enters in second
order in the low-energy expansion: $\epsilon({\pmb\kappa})=\pm
v_f|{\pmb\kappa}|+v_f^2|{\pmb\kappa}|^2[t^{\prime}/t^2\mp\nu\cos(3\beta)/(6t)]+...$
These contributions are only important far from the Fermi level and we
neglect them.

The linear response is defined
as\cite{MahanBook}
\begin{equation}
\left(
\begin{array}{c}
{\bf j}\\
{\bf j}_E
\end{array}
\right) = \left(
\begin{array}{cc}
{\cal L}_{11} & {\cal L}_{12}\\
{\cal L}_{21} & {\cal L}_{22}
\end{array}
\right) \left(
\begin{array}{c}
\frac{\bf E}{T}\\
\nabla\frac{1}{T}
\end{array}
\right),
\end{equation}
where ${\bf E}$ is the electric field and $T$ is the temperature. The
Onsager relation ${\cal L}_{12}={\cal L}_{21}$ holds. The charge
conductivity is defined as $\sigma=e^2{\cal L}_{11}/T$, while the
electronic contribution to the open-circuit heat conductivity is
$\kappa^{el}=(1/T)({\cal L}_{22}-{\cal L}_{12}^2/{\cal
  L}_{11})$. Finally, the thermopower is defined as $S=-{\cal
  L}_{12}/(eT{\cal L}_{11})$. The response functions ${\cal L}_{ij}$
are defined in terms of Kubo formulas, i.e. current-current
correlation functions. The charge current operator for graphene
modeled as above has the simple form
$
{\bf j}=\sum_{\bf k} \left[{\bf v}_{\bf k} a_{\bf k}^{\dagger}b_{\bf k} + h.c.\right],
$
where $h.c.$ denotes hermitian conjugate, and ${\bf v}_{\bf
  k}=\nabla_{\bf k}\phi_{\bf k}$. Note that the function $\phi_{\bf
  k}=-t\sum_{j=1}^3e^{i{\bf k}\cdot{\pmb\delta}_j}$, and therefore
also ${\bf v}_{\bf k}$, are complex functions since the three nearest
neighbor vectors ${\pmb\delta}_j$ point in three directions rotated
$120^o$ relative to each other in the graphene plane. The heat current
operator has an extra term from impurity scattering,
${\bf j}_E={\bf j}_E^0+{\bf j}_E^{\sf imp}$, where
%
$
{\bf j}_E^0 = (1/2)\sum_{\bf k}
({\bf v}_{\bf k}\phi_{\bf k}^*+{\bf v}_{\bf k}^*\phi_{\bf k})
(a_{\bf k}^{\dagger}a_{\bf k}+b_{\bf k}^{\dagger}b_{\bf k}),
$
and
$
{\bf j}_E^{\sf imp} (1/2)\sum_{{\bf kq}}\left[
(U_{\bf q}^B{\bf v}_{\bf k}+U_{\bf q}^A{\bf v}_{{\bf k}+{\bf q}})
a_{\bf k}^{\dagger}b_{{\bf k}+{\bf q}}+h.c.\right]
$,
%
which in principle complicates the calculation of the thermal
response. However, for graphene, as was shown for normal metals by
Jonson and Mahan\cite{jon80,jon90} (see also
Refs.~\onlinecite{pau03}), the response function kernels are simply
related to each other: once we know the charge current response kernel
$K_{11}=K(\epsilon)$ we get the other two kernels through
$K_{12}=\epsilon K(\epsilon)$ and $K_{22}=\epsilon^2 K(\epsilon)$. The
repsonse functions are then computed by integration
\begin{equation}
{\cal L}_{ij}=\frac{1}{\pi^2}\int_{-\infty}^{\infty} d\epsilon
\left(
-\frac{\partial f(\epsilon)}{\partial\epsilon} 
\right) K_{ij}(\epsilon),
\label{eq:Lij}
\end{equation}
where $f(\epsilon)$ is the Fermi distribution function. For point
scatterers the kernel $K(\epsilon)$ can be approximated by the bare
bubble.\cite{per06,ost06} The needed sum over ${\bf k}$ can be
computed analytically in the K-point approximation. We write
$z=\epsilon-\Sigma^R(\epsilon)=a(\epsilon)+ib(\epsilon)$ and
find\cite{MahanBook,per06}
\begin{eqnarray}
\label{eq:kernel}
&&K(\epsilon) = 1 + \frac{a^2+b^2}{ab}\arctan\frac{a}{b}\\
&&- \sum _{s=\pm 1} s\left[
\frac{a(\epsilon_c+sa)+sb^2}{(\epsilon_c+sa)^2+b^2}
- \frac{a^2+b^2}{2ab}\arctan\left( \frac{\epsilon_c+sa}{b} \right) \right]
\nonumber
\end{eqnarray}

\begin{figure}[t]
\includegraphics[width=\columnwidth]{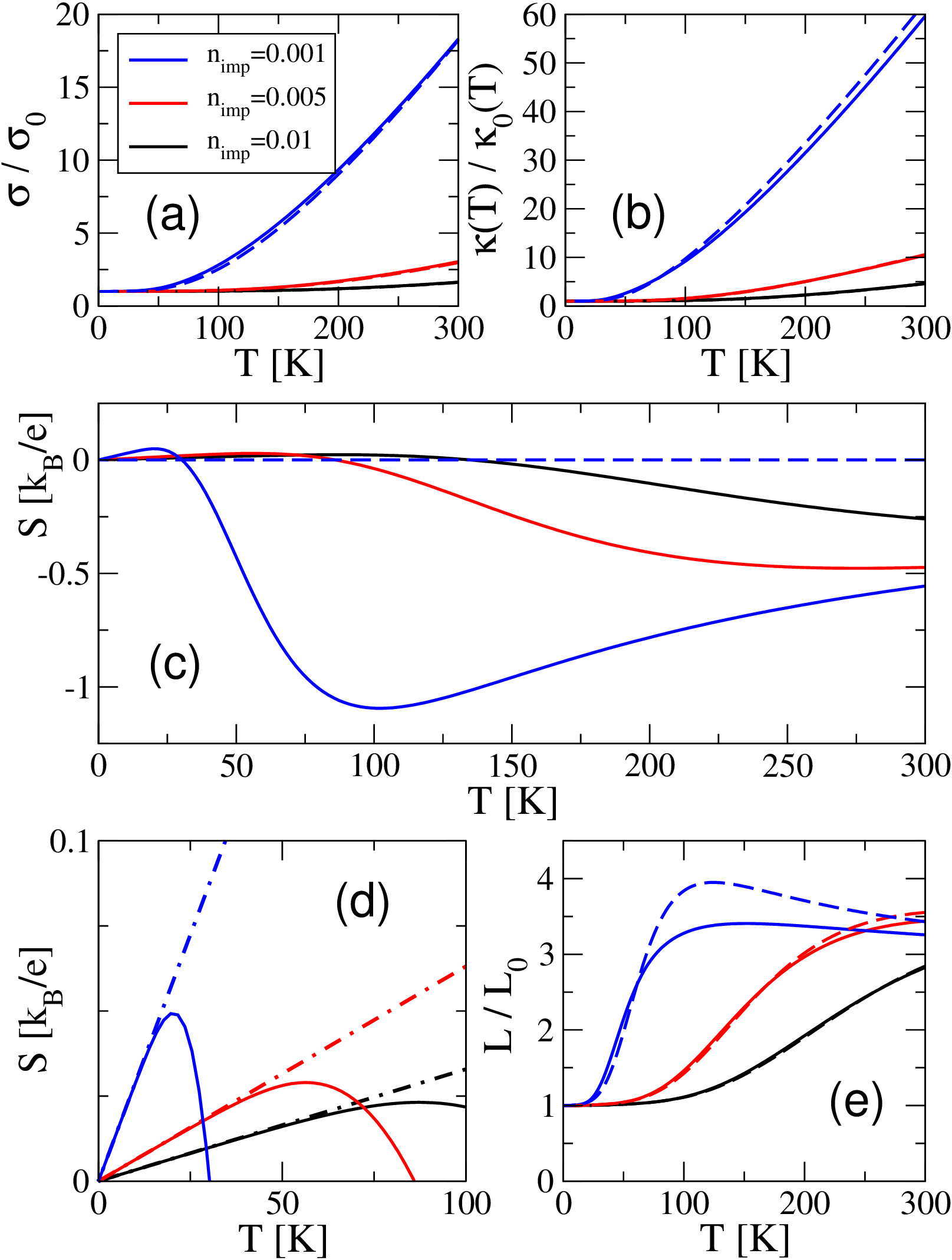}
\caption{Temperature dependence of (a) the charge conductance, (b) the
  heat conductance, and (c) the thermopower for three values of the
  impurity density $n_{\sf imp}$. Solid lines are for a large impurity
  potential $V_{\sf imp}=20\epsilon_c$ and the dashed lines for the strict
  unitarly limit $V_{\sf imp}\rightarrow\infty$. In (a)-(b) we have
  normalized the conductivities by the low-temperature asymptotics
  $\sigma_0=4e^2/(\pi h)$ and $\kappa_0(T)=4\pi k_B^2T/(3h)$. In (d)
  we show the details at low temperatures. The dash-dotted lines are
  the thermopower computed throught the Mott relation. In (e) we show
  the Lorenz ration $L=\kappa/(\sigma T)$ in units of the value
  $L_0=(\pi^2/3)(k_B^2/e^2)$ appearing in Wiedeman-Franz law. We used
  $\epsilon_c=1eV\rightarrow 11605K$ to convert the temperature scale
  to Kelvin.}
\label{fig:conduct}
\end{figure}

In Fig.~\ref{fig:conduct} we present results for conductivities and
the thermopower. The conductivities are only weakly dependent on the
exact value of $V_{\sf imp}$ as long as $V_{\sf imp}\gg\epsilon_c$, as
is clear when we compare the solid and dashed lines in
Fig.~\ref{fig:conduct}(a)-(b). These response functions are given by
the electron-hole symmetric part of $K(\epsilon)$ which is robust as
long as the impurity band is not shifted far from the Dirac point. We
conclude that, for these response functions, the unitary limit is
effectively approached quickly for $V_{\sf imp}\gg\epsilon_c$. On the
other hand, the thermopower is given by the electron-hole asymmetric
part of $K(\epsilon)$ and is very sensitive to impurity scattering. We
have $S(-V_{\sf imp})=-S(V_{\sf imp})$. The order of magnitude $S\sim
k_B/e\sim 100\mu V/K$ is in agreement with experiments on single-wall
carbon nanotubes.\cite{sma03}

At low temperatures both the charge conductivity and the slope of the
thermal conductivity reaches constant universal values
$\sigma_0=4e^2/(\pi h)$ and $\kappa_0=4\pi k_B^2T/(3h)$, respectively,
that are independent of the details of the impurities, in agreement
with results in the literature.\cite{sho98,gus05b,gus06,per06,ost06}
On the other hand, the low-temperature slope of the thermopower
reaches a non-universal constant value, that depends sensitively on
the nature of impurity scattering. These results are understood in
terms of a Sommerfeld expansions of the transport coefficitents in the
small parameter $T/\gamma\ll 1$ for slowly varying kernels
$K_{ij}(\epsilon)=K_{ij}(0)+\epsilon
\left.K_{ij}^{\prime}(\epsilon)\right|_{\epsilon=0}+...$. The needed
parts of the kernel Eq.~(\ref{eq:kernel}) can be computed
analytically, $K(0) = 2$, and $K^{\prime}(0) =
4\alpha(1-\alpha)\epsilon_r/(3\gamma^2)$, where we assumed that
$\epsilon_r\ll\gamma\ll\epsilon_c$. The universal conductivities
immediately follows, and the low-temperature thermopower is
(reinstating $\hbar$ and $k_B$)
\begin{equation}
S \simeq
-\frac{\pi^2}{3}\frac{2\alpha(1-\alpha)\epsilon_r}{3\gamma^2}\frac{k_B^2T}{e}.
\label{eq:S}
\end{equation}
This means that at low temperatures, $k_BT\ll\gamma$, the
Wiedemann-Franz law is obeyed, $L=\kappa^{el}/\sigma
T=L_0=(\pi^2/3)(k_B^2/e^2)$, and also the Mott formula holds,
$S=-(\pi^2/3)(k_B^2T/e)\left.d[\ln
  K(\epsilon)]/d\epsilon\right|_{\epsilon\rightarrow 0}$. At higher
temperatures $T\agt\gamma$, however, neither of these relations hold,
see Fig.~\ref{fig:conduct}(d)-(e). This happens for any system where
the conductivity kernel is varying around the Fermi level on some
particular energy scale, here given by the impurity band width
$\gamma$. For normal metals this scale is typically given by the much
larger Fermi energy.

There are other contributions to the heat conductance besides the
electronic. In particular phonons are important in graphite and carbon
nanotubes,\cite{dre00R} for which the Lorenz ration is typically found
to be $L\sim 10L_0-100L_0$.

In contrast, the thermopower is given entirely by electronic
contributions. In the low-temperature limit, $T\ll\gamma$, we see that
$S$ is proportional to $\epsilon_r$ and $\gamma^{-2}$, where for large
$V_{\sf imp}\gg\epsilon_c$, $\epsilon_r\propto \epsilon_c^2/V_{\sf imp}$ and
$\gamma\propto \epsilon_c\sqrt{n_{\sf imp}}$. The thermopower is therefore
propotional to the inverses $1/V_{\sf imp}$ and $1/n_{\sf imp}$. In the strict
unitary limit, $V_{\sf imp}\rightarrow\infty$, when electron-hole symmetry
is restored, the thermopower vanishes.\cite{gus06,dor07} But for a
large (but not infinitely large) impurity potential the thermopower is
enhanced: the smallness of $1/V_{\sf imp}$ is compensated by the large
$1/n_{\sf imp}$. This also means that if $n_{\sf imp}$ can be controlled,
$V_{\sf imp}$ can be extracted by measuring the slope of the thermopower
at low temperatures.

We note that there are clear analogies with the situation in a
$d$-wave superconductor,\cite{graf96} where the Lorenz ratio deviates
from $L_0$ but recovers at low temperatures, $T\ll\gamma$. An
anomalously large thermoelectric coefficient ${\cal L}_{12}$ has been
predicted.\cite{LF04,LF05} A significant difference for graphene
compared with the superconducting case is the possibility to measure
the thermoelectric response directly through the thermopower, which is
not possible in a superconductor since supercurrents short-circuit the
thermoelectric voltage.

\begin{figure}[t]
\includegraphics[width=\columnwidth]{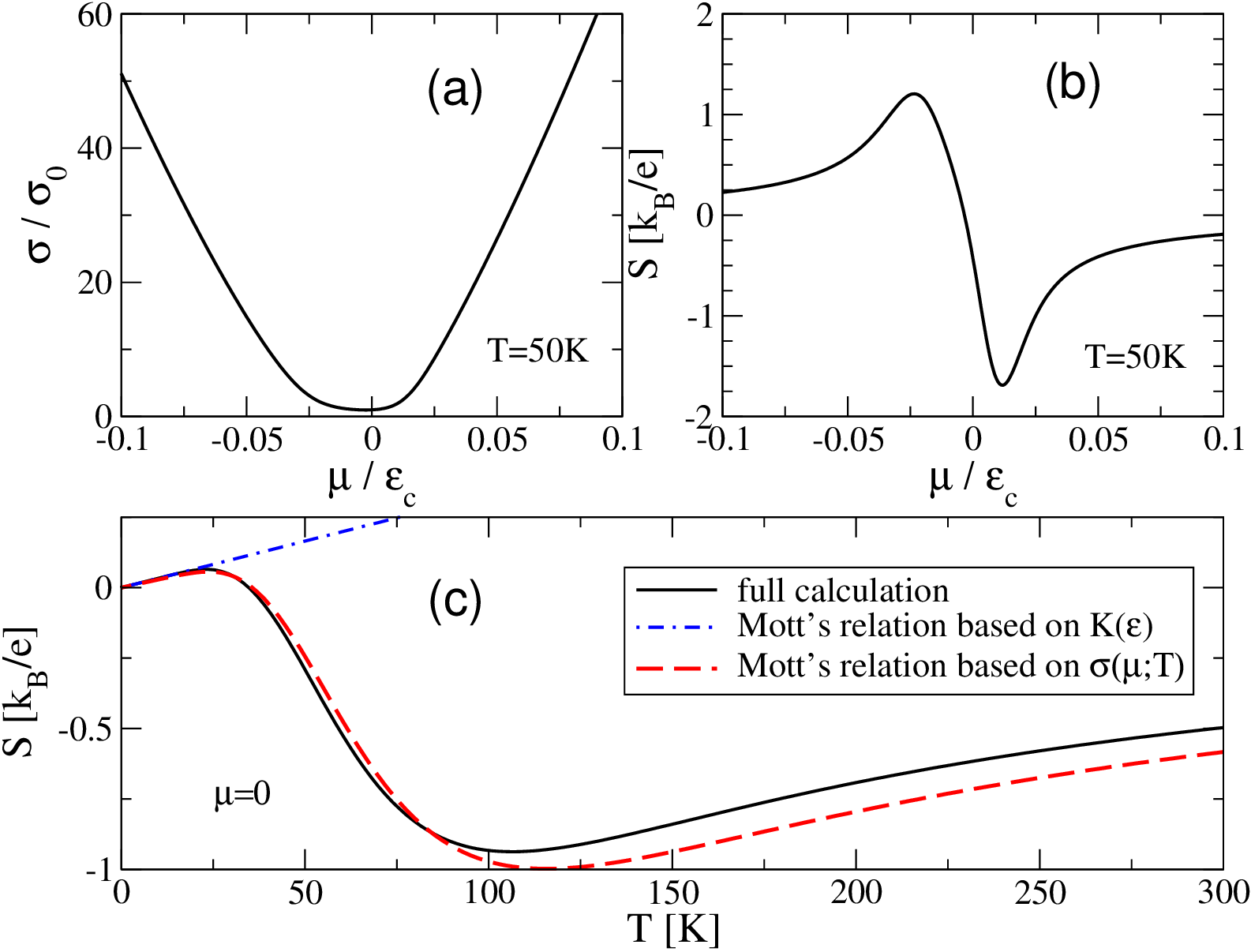}
\caption{The charge conductance (a) and the thermopower (b) as
  function of chemical potential at $T=50K$. In (c) we show how thermal
  smearing leads to an approximate Mott's relation based on the charge
  conductance as function of chemical potential. For $T\ll\gamma$, the
  Mott relation is exact to leading order (dash-dotted line).}
\label{fig:EffMott}
\end{figure}

The chemical potential of graphene can be tuned by applying a voltage
to the substrate.\cite{nov04} We show the dependence on the chemical
potential in Fig.~\ref{fig:EffMott}(a)-(b). The charge conductance as
function of $\mu$ is essentially linear at large $\mu$. The value of
the thermpower at $\mu=0$ and the associated asymmetry of $S$ around
$\mu=0$, and also around the point $\mu^*$, where $S(\mu^*)=0$, is
related to the amount of electron-hole asymmetry caused by impurity
scattering.

Mott's formula, derived mathematically through the Sommerfeld
expansion, is only valid at low temperatures $T\ll\gamma$, since the
conductivity kernel $K(\epsilon)$ varies slowly only on the energy
scale $\gamma$. But as we show in Fig.~\ref{fig:EffMott}(c), it turns
out that thermal smearing leads to an effective, approximate Mott
relation $S\approx-(\pi^2/3)(k_B^2T/e)d[\ln\sigma(\mu;T)]/d\mu$ in
terms of the full temperature dependent conductivity. The non-linear
temperature dependence $S(T)$ is obtained at $\mu=0$.

In summary, we have presented results for the linear response to
electric and thermal forces in graphene for the case of strong
impurity scattering, near the unitary limit. The impurity band of
width $\gamma$ is centered near the Fermi level, at an energy
$\epsilon_r\propto\epsilon_c^2/V_{\sf imp}$. The induced electron-hole
asymmetry gives small changes of the charge and thermal
conductivities, but leads to an enhanced thermopower, which at low
temperatures $T\ll\gamma$ is linear in temperature with a slope
proportional to $1/(V_{\sf imp}n_{\sf imp})$. The thermopower,
measured directly or estimated by Mott's relation, can therefore be
used to extract information about impurity scattering in graphene.

{\bf Acknowledgments.}  It is a pleasure to thank A. Yurgens and
V. Shumeiko for valuable discussions. Financial support from SSF, the
Swedish Foundation for Strategic Reserach (T.L.), and the Swedish
Research Council (M.F.), is gratefully acknowledged.


\begin{thebibliography}{40}
\expandafter\ifx\csname natexlab\endcsname\relax\def\natexlab#1{#1}\fi
\expandafter\ifx\csname bibnamefont\endcsname\relax
  \def\bibnamefont#1{#1}\fi
\expandafter\ifx\csname bibfnamefont\endcsname\relax
  \def\bibfnamefont#1{#1}\fi
\expandafter\ifx\csname citenamefont\endcsname\relax
  \def\citenamefont#1{#1}\fi
\expandafter\ifx\csname url\endcsname\relax
  \def\url#1{\texttt{#1}}\fi
\expandafter\ifx\csname urlprefix\endcsname\relax\def\urlprefix{URL }\fi
\providecommand{\bibinfo}[2]{#2}
\providecommand{\eprint}[2][]{\url{#2}}

\bibitem[{\citenamefont{Novoselov et~al.}(2004)\citenamefont{Novoselov, Geim,
  Morozov, Jiang, Zhang, Dubonos, Grigorieva, and Firsov}}]{nov04}
\bibinfo{author}{\bibfnamefont{K.~S.} \bibnamefont{Novoselov}},
  \bibinfo{author}{\bibfnamefont{A.~K.} \bibnamefont{Geim}},
  \bibinfo{author}{\bibfnamefont{S.~V.} \bibnamefont{Morozov}},
  \bibinfo{author}{\bibfnamefont{D.}~\bibnamefont{Jiang}},
  \bibinfo{author}{\bibfnamefont{Y.}~\bibnamefont{Zhang}},
  \bibinfo{author}{\bibfnamefont{S.~V.} \bibnamefont{Dubonos}},
  \bibinfo{author}{\bibfnamefont{I.~V.} \bibnamefont{Grigorieva}},
  \bibnamefont{and} \bibinfo{author}{\bibfnamefont{A.~A.}
  \bibnamefont{Firsov}}, \bibinfo{journal}{Science}
  \textbf{\bibinfo{volume}{306}}, \bibinfo{pages}{666} (\bibinfo{year}{2004}).

\bibitem[{\citenamefont{Novoselov et~al.}(2005)\citenamefont{Novoselov, Geim,
  Morozov, Jiang, Katsnelson, Grigorieva, Dubonos, and Firsov}}]{nov05b}
\bibinfo{author}{\bibfnamefont{K.~S.} \bibnamefont{Novoselov}},
  \bibinfo{author}{\bibfnamefont{A.~K.} \bibnamefont{Geim}},
  \bibinfo{author}{\bibfnamefont{S.~V.} \bibnamefont{Morozov}},
  \bibinfo{author}{\bibfnamefont{D.}~\bibnamefont{Jiang}},
  \bibinfo{author}{\bibfnamefont{M.~I.} \bibnamefont{Katsnelson}},
  \bibinfo{author}{\bibfnamefont{I.~V.} \bibnamefont{Grigorieva}},
  \bibinfo{author}{\bibfnamefont{S.~V.} \bibnamefont{Dubonos}},
  \bibnamefont{and} \bibinfo{author}{\bibfnamefont{A.~A.}
  \bibnamefont{Firsov}}, \bibinfo{journal}{Nature}
  \textbf{\bibinfo{volume}{438}}, \bibinfo{pages}{197} (\bibinfo{year}{2005}).

\bibitem[{\citenamefont{Zhang et~al.}(2005)\citenamefont{Zhang, Tan, Stormer,
  and Kim}}]{zha05}
\bibinfo{author}{\bibfnamefont{Y.}~\bibnamefont{Zhang}},
  \bibinfo{author}{\bibfnamefont{Y.-W.} \bibnamefont{Tan}},
  \bibinfo{author}{\bibfnamefont{H.~L.} \bibnamefont{Stormer}},
  \bibnamefont{and} \bibinfo{author}{\bibfnamefont{P.}~\bibnamefont{Kim}},
  \bibinfo{journal}{Nature} \textbf{\bibinfo{volume}{438}},
  \bibinfo{pages}{201} (\bibinfo{year}{2005}).

\bibitem[{\citenamefont{Geim and Novoselov}(2007)}]{geim07R}
\bibinfo{author}{\bibfnamefont{A.~K.} \bibnamefont{Geim}} \bibnamefont{and}
  \bibinfo{author}{\bibfnamefont{K.~S.} \bibnamefont{Novoselov}},
  \bibinfo{journal}{Nature Materials} \textbf{\bibinfo{volume}{6}},
  \bibinfo{pages}{183} (\bibinfo{year}{2007}).

\bibitem[{\citenamefont{Berger et~al.}(2006)\citenamefont{Berger, Song, Li, Wu,
  Brown, Naud, Mayou, Li, Hass, Marchenkov et~al.}}]{ber06}
\bibinfo{author}{\bibfnamefont{C.}~\bibnamefont{Berger}},
  \bibinfo{author}{\bibfnamefont{Z.}~\bibnamefont{Song}},
  \bibinfo{author}{\bibfnamefont{X.}~\bibnamefont{Li}},
  \bibinfo{author}{\bibfnamefont{X.}~\bibnamefont{Wu}},
  \bibinfo{author}{\bibfnamefont{N.}~\bibnamefont{Brown}},
  \bibinfo{author}{\bibfnamefont{C.}~\bibnamefont{Naud}},
  \bibinfo{author}{\bibfnamefont{D.}~\bibnamefont{Mayou}},
  \bibinfo{author}{\bibfnamefont{T.}~\bibnamefont{Li}},
  \bibinfo{author}{\bibfnamefont{J.}~\bibnamefont{Hass}},
  \bibinfo{author}{\bibfnamefont{A.~N.} \bibnamefont{Marchenkov}},
  \bibnamefont{et~al.}, \bibinfo{journal}{Science}
  \textbf{\bibinfo{volume}{312}}, \bibinfo{pages}{1191} (\bibinfo{year}{2006}).

\bibitem[{\citenamefont{Ferrari et~al.}(2006)\citenamefont{Ferrari, Meyer,
  Scardaci, Casiraghi, Lazzeri, Mauri, Piscanec, Jiang, Novoselov, Roth
  et~al.}}]{fer06}
\bibinfo{author}{\bibfnamefont{A.~C.} \bibnamefont{Ferrari}},
  \bibinfo{author}{\bibfnamefont{J.~C.} \bibnamefont{Meyer}},
  \bibinfo{author}{\bibfnamefont{V.}~\bibnamefont{Scardaci}},
  \bibinfo{author}{\bibfnamefont{C.}~\bibnamefont{Casiraghi}},
  \bibinfo{author}{\bibfnamefont{M.}~\bibnamefont{Lazzeri}},
  \bibinfo{author}{\bibfnamefont{F.}~\bibnamefont{Mauri}},
  \bibinfo{author}{\bibfnamefont{S.}~\bibnamefont{Piscanec}},
  \bibinfo{author}{\bibfnamefont{D.}~\bibnamefont{Jiang}},
  \bibinfo{author}{\bibfnamefont{K.~S.} \bibnamefont{Novoselov}},
  \bibinfo{author}{\bibfnamefont{S.}~\bibnamefont{Roth}}, \bibnamefont{et~al.},
  \bibinfo{journal}{Phys. Rev. Lett.} \textbf{\bibinfo{volume}{97}},
  \bibinfo{pages}{187401} (\bibinfo{year}{2006}).

\bibitem[{\citenamefont{Heersche et~al.}(2007)\citenamefont{Heersche,
  Jarillo-Herrero, Oostinga, Vandersypen, and Morpurgo}}]{hee07}
\bibinfo{author}{\bibfnamefont{H.~B.} \bibnamefont{Heersche}},
  \bibinfo{author}{\bibfnamefont{P.}~\bibnamefont{Jarillo-Herrero}},
  \bibinfo{author}{\bibfnamefont{J.~B.} \bibnamefont{Oostinga}},
  \bibinfo{author}{\bibfnamefont{L.~M.~K.} \bibnamefont{Vandersypen}},
  \bibnamefont{and} \bibinfo{author}{\bibfnamefont{A.~F.}
  \bibnamefont{Morpurgo}}, \bibinfo{journal}{Nature}
  \textbf{\bibinfo{volume}{446}}, \bibinfo{pages}{56} (\bibinfo{year}{2007}).

\bibitem[{\citenamefont{Bostwick et~al.}(2007)\citenamefont{Bostwick, Ohta,
  Seyller, Horn, and Rotenberg}}]{bos07}
\bibinfo{author}{\bibfnamefont{A.}~\bibnamefont{Bostwick}},
  \bibinfo{author}{\bibfnamefont{T.}~\bibnamefont{Ohta}},
  \bibinfo{author}{\bibfnamefont{T.}~\bibnamefont{Seyller}},
  \bibinfo{author}{\bibfnamefont{K.}~\bibnamefont{Horn}}, \bibnamefont{and}
  \bibinfo{author}{\bibfnamefont{E.}~\bibnamefont{Rotenberg}},
  \bibinfo{journal}{Nature Physics} \textbf{\bibinfo{volume}{3}},
  \bibinfo{pages}{36} (\bibinfo{year}{2007}).

\bibitem[{\citenamefont{Schedin et~al.}()\citenamefont{Schedin, Novoselov,
  Morozov, Jiang, Hill, Blake, and Geim}}]{sch07}
\bibinfo{author}{\bibfnamefont{F.}~\bibnamefont{Schedin}},
  \bibinfo{author}{\bibfnamefont{K.~S.} \bibnamefont{Novoselov}},
  \bibinfo{author}{\bibfnamefont{S.~V.} \bibnamefont{Morozov}},
  \bibinfo{author}{\bibfnamefont{D.}~\bibnamefont{Jiang}},
  \bibinfo{author}{\bibfnamefont{E.~H.} \bibnamefont{Hill}},
  \bibinfo{author}{\bibfnamefont{P.}~\bibnamefont{Blake}}, \bibnamefont{and}
  \bibinfo{author}{\bibfnamefont{A.~K.} \bibnamefont{Geim}},
  \bibinfo{note}{cond-mat/0610809}.

\bibitem[{\citenamefont{Chen et~al.}()\citenamefont{Chen, Lin, Rooks, and
  Avouris}}]{chen07}
\bibinfo{author}{\bibfnamefont{Z.}~\bibnamefont{Chen}},
  \bibinfo{author}{\bibfnamefont{Y.-M.} \bibnamefont{Lin}},
  \bibinfo{author}{\bibfnamefont{M.~J.} \bibnamefont{Rooks}}, \bibnamefont{and}
  \bibinfo{author}{\bibfnamefont{P.}~\bibnamefont{Avouris}},
  \bibinfo{note}{cond-mat/0701599}.

\bibitem[{\citenamefont{Trauzettel et~al.}(2007)\citenamefont{Trauzettel,
  Bulaev, Loss, and Burkard}}]{tra07}
\bibinfo{author}{\bibfnamefont{B.}~\bibnamefont{Trauzettel}},
  \bibinfo{author}{\bibfnamefont{D.~V.} \bibnamefont{Bulaev}},
  \bibinfo{author}{\bibfnamefont{D.}~\bibnamefont{Loss}}, \bibnamefont{and}
  \bibinfo{author}{\bibfnamefont{G.}~\bibnamefont{Burkard}},
  \bibinfo{journal}{Nature Physics} \textbf{\bibinfo{volume}{3}},
  \bibinfo{pages}{192} (\bibinfo{year}{2007}).

\bibitem[{\citenamefont{Rycerz et~al.}(2007)\citenamefont{Rycerz, Tworzydlo,
  and Beenakker}}]{ryc07}
\bibinfo{author}{\bibfnamefont{A.}~\bibnamefont{Rycerz}},
  \bibinfo{author}{\bibfnamefont{J.}~\bibnamefont{Tworzydlo}},
  \bibnamefont{and} \bibinfo{author}{\bibfnamefont{C.~W.~J.}
  \bibnamefont{Beenakker}}, \bibinfo{journal}{Nature Physics}
  \textbf{\bibinfo{volume}{3}}, \bibinfo{pages}{172} (\bibinfo{year}{2007}).

\bibitem[{\citenamefont{Cheianov et~al.}(2007)\citenamefont{Cheianov, {Fal'ko},
  and Altshuler}}]{che07}
\bibinfo{author}{\bibfnamefont{V.~V.} \bibnamefont{Cheianov}},
  \bibinfo{author}{\bibfnamefont{V.}~\bibnamefont{{Fal'ko}}}, \bibnamefont{and}
  \bibinfo{author}{\bibfnamefont{B.~L.} \bibnamefont{Altshuler}},
  \bibinfo{journal}{Science} \textbf{\bibinfo{volume}{315}},
  \bibinfo{pages}{1252} (\bibinfo{year}{2007}).

\bibitem[{\citenamefont{Son et~al.}(2006)\citenamefont{Son, Cohen, and
  Louie}}]{son06}
\bibinfo{author}{\bibfnamefont{Y.-W.} \bibnamefont{Son}},
  \bibinfo{author}{\bibfnamefont{M.~L.} \bibnamefont{Cohen}}, \bibnamefont{and}
  \bibinfo{author}{\bibfnamefont{S.~G.} \bibnamefont{Louie}},
  \bibinfo{journal}{Nature} \textbf{\bibinfo{volume}{444}},
  \bibinfo{pages}{347} (\bibinfo{year}{2006}).

\bibitem[{\citenamefont{Silvestrov and Efetov}(2007)}]{sil07}
\bibinfo{author}{\bibfnamefont{P.~G.} \bibnamefont{Silvestrov}}
  \bibnamefont{and} \bibinfo{author}{\bibfnamefont{K.~B.}
  \bibnamefont{Efetov}}, \bibinfo{journal}{Phys. Rev. Lett.}
  \textbf{\bibinfo{volume}{98}}, \bibinfo{pages}{016802}
  (\bibinfo{year}{2007}).

\bibitem[{\citenamefont{Katsnelson}(2007)}]{kat07R}
\bibinfo{author}{\bibfnamefont{M.~I.} \bibnamefont{Katsnelson}},
  \bibinfo{journal}{Materials Today}
  \bibinfo{volume}{\textbf{10}, issues 1-2},
  \bibinfo{pages}{20} (\bibinfo{year}{2007}).

\bibitem[{\citenamefont{Wallace}(1947)}]{wal47}
\bibinfo{author}{\bibfnamefont{P.~R.} \bibnamefont{Wallace}},
  \bibinfo{journal}{Phys. Rev.} \textbf{\bibinfo{volume}{71}},
  \bibinfo{pages}{622} (\bibinfo{year}{1947}).

\bibitem[{\citenamefont{Slonczewski and Weiss}(1958)}]{slo58}
\bibinfo{author}{\bibfnamefont{J.~C.} \bibnamefont{Slonczewski}}
  \bibnamefont{and} \bibinfo{author}{\bibfnamefont{P.~R.} \bibnamefont{Weiss}},
  \bibinfo{journal}{Phys. Rev.} \textbf{\bibinfo{volume}{109}},
  \bibinfo{pages}{272} (\bibinfo{year}{1958}).

\bibitem[{\citenamefont{Shon and Ando}(1998)}]{sho98}
\bibinfo{author}{\bibfnamefont{N.~H.} \bibnamefont{Shon}} \bibnamefont{and}
  \bibinfo{author}{\bibfnamefont{T.}~\bibnamefont{Ando}}, \bibinfo{journal}{J.
  Phys. Soc. Jpn.} \textbf{\bibinfo{volume}{67}}, \bibinfo{pages}{2421}
  (\bibinfo{year}{1998}).

\bibitem[{\citenamefont{Gusynin and Sharapov}(2005)}]{gus05b}
\bibinfo{author}{\bibfnamefont{V.~P.} \bibnamefont{Gusynin}} \bibnamefont{and}
  \bibinfo{author}{\bibfnamefont{S.~G.} \bibnamefont{Sharapov}},
  \bibinfo{journal}{Phys. Rev. Lett.} \textbf{\bibinfo{volume}{95}},
  \bibinfo{pages}{146801} (\bibinfo{year}{2005}).

\bibitem[{\citenamefont{Gusynin and Sharapov}(2006)}]{gus06}
\bibinfo{author}{\bibfnamefont{V.~P.} \bibnamefont{Gusynin}} \bibnamefont{and}
  \bibinfo{author}{\bibfnamefont{S.~G.} \bibnamefont{Sharapov}},
  \bibinfo{journal}{Phys. Rev. B} \textbf{\bibinfo{volume}{73}},
  \bibinfo{pages}{245411} (\bibinfo{year}{2006}).

\bibitem[{\citenamefont{Peres et~al.}(2006)\citenamefont{Peres, Guinea, and
  {Casto Neto}}}]{per06}
\bibinfo{author}{\bibfnamefont{N.~M.~R.} \bibnamefont{Peres}},
  \bibinfo{author}{\bibfnamefont{F.}~\bibnamefont{Guinea}}, \bibnamefont{and}
  \bibinfo{author}{\bibfnamefont{A.~H.} \bibnamefont{{Casto Neto}}},
  \bibinfo{journal}{Phys. Rev. B} \textbf{\bibinfo{volume}{73}},
  \bibinfo{pages}{125411} (\bibinfo{year}{2006}).

\bibitem[{\citenamefont{Ostrovsky et~al.}(2006)\citenamefont{Ostrovsky, Gornyi,
  and Mirlin}}]{ost06}
\bibinfo{author}{\bibfnamefont{P.~M.} \bibnamefont{Ostrovsky}},
  \bibinfo{author}{\bibfnamefont{I.~V.} \bibnamefont{Gornyi}},
  \bibnamefont{and} \bibinfo{author}{\bibfnamefont{A.~D.}
  \bibnamefont{Mirlin}}, \bibinfo{journal}{Phys. Rev. B}
  \textbf{\bibinfo{volume}{74}}, \bibinfo{pages}{235443}
  (\bibinfo{year}{2006}).

\bibitem[{\citenamefont{Ziegler}(2006)}]{zie06}
\bibinfo{author}{\bibfnamefont{K.}~\bibnamefont{Ziegler}},
  \bibinfo{journal}{Phys. Rev. Lett.} \textbf{\bibinfo{volume}{97}},
  \bibinfo{pages}{266802} (\bibinfo{year}{2006}).

\bibitem[{\citenamefont{Nomura and MacDonald}(2007)}]{nom07}
\bibinfo{author}{\bibfnamefont{K.}~\bibnamefont{Nomura}} \bibnamefont{and}
  \bibinfo{author}{\bibfnamefont{A.~H.} \bibnamefont{MacDonald}},
  \bibinfo{journal}{Phys. Rev. Lett.} \textbf{\bibinfo{volume}{98}},
  \bibinfo{pages}{076602} (\bibinfo{year}{2007}).

\bibitem[{\citenamefont{{Das Sarma} et~al.}(2007)\citenamefont{{Das Sarma},
  Hwang, and Tse}}]{das07}
\bibinfo{author}{\bibfnamefont{S.}~\bibnamefont{{Das Sarma}}},
  \bibinfo{author}{\bibfnamefont{E.~H.} \bibnamefont{Hwang}}, \bibnamefont{and}
  \bibinfo{author}{\bibfnamefont{W.-K.} \bibnamefont{Tse}},
  \bibinfo{journal}{Phys. Rev. B} \textbf{\bibinfo{volume}{75}},
  \bibinfo{pages}{121406} (\bibinfo{year}{2007}).

\bibitem[{\citenamefont{Ando}(2005)}]{ando05R}
\bibinfo{author}{\bibfnamefont{T.}~\bibnamefont{Ando}}, \bibinfo{journal}{J.
  Phys. Soc. Jpn.} \textbf{\bibinfo{volume}{74}}, \bibinfo{pages}{777}
  (\bibinfo{year}{2005}).

\bibitem[{\citenamefont{Mahan}(2004)}]{mah04}
\bibinfo{author}{\bibfnamefont{G.~D.} \bibnamefont{Mahan}},
  \bibinfo{journal}{Phys. Rev. B} \textbf{\bibinfo{volume}{69}},
  \bibinfo{pages}{125407} (\bibinfo{year}{2004}).

\bibitem[{\citenamefont{Pogorelov}()}]{pog06}
\bibinfo{author}{\bibfnamefont{Y.~G.} \bibnamefont{Pogorelov}},
  \bibinfo{note}{cond-mat/0603327}.

\bibitem[{\citenamefont{Wehling et~al.}(2007)\citenamefont{Wehling, Balatsky,
  Katsnelson, Lichtenstein, Scharnberg, and Wiesendanger}}]{weh07}
\bibinfo{author}{\bibfnamefont{T.~O.} \bibnamefont{Wehling}},
  \bibinfo{author}{\bibfnamefont{A.~V.} \bibnamefont{Balatsky}},
  \bibinfo{author}{\bibfnamefont{M.~I.} \bibnamefont{Katsnelson}},
  \bibinfo{author}{\bibfnamefont{A.~I.} \bibnamefont{Lichtenstein}},
  \bibinfo{author}{\bibfnamefont{K.}~\bibnamefont{Scharnberg}},
  \bibnamefont{and}
  \bibinfo{author}{\bibfnamefont{R.}~\bibnamefont{Wiesendanger}},
  \bibinfo{journal}{Phys. Rev. B} \textbf{\bibinfo{volume}{75}},
  \bibinfo{pages}{125425} (\bibinfo{year}{2007}).

\bibitem[{\citenamefont{{D\'ora} and Thalmeier}()}]{dor07}
\bibinfo{author}{\bibfnamefont{B.}~\bibnamefont{{D\'ora}}} \bibnamefont{and}
  \bibinfo{author}{\bibfnamefont{P.}~\bibnamefont{Thalmeier}},
  \bibinfo{note}{cond-mat/0701714}.

\bibitem[{\citenamefont{Mahan}(1990)}]{MahanBook}
\bibinfo{author}{\bibfnamefont{G.~D.} \bibnamefont{Mahan}},
  \emph{\bibinfo{title}{Many-Particle Physics}}, \bibinfo{number}{2nd ed.}
  (\bibinfo{publisher}{Plenum Press}, \bibinfo{address}{New York},
  \bibinfo{year}{1990}).

\bibitem[{\citenamefont{Jonson and Mahan}(1980)}]{jon80}
\bibinfo{author}{\bibfnamefont{M.}~\bibnamefont{Jonson}} \bibnamefont{and}
  \bibinfo{author}{\bibfnamefont{G.~D.} \bibnamefont{Mahan}},
  \bibinfo{journal}{Phys. Rev. B} \textbf{\bibinfo{volume}{21}},
  \bibinfo{pages}{4223} (\bibinfo{year}{1980}).

\bibitem[{\citenamefont{Jonson and Mahan}(1990)}]{jon90}
\bibinfo{author}{\bibfnamefont{M.}~\bibnamefont{Jonson}} \bibnamefont{and}
  \bibinfo{author}{\bibfnamefont{G.~D.} \bibnamefont{Mahan}},
  \bibinfo{journal}{Phys. Rev. B} \textbf{\bibinfo{volume}{42}},
  \bibinfo{pages}{9350} (\bibinfo{year}{1990}).

\bibitem[{\citenamefont{Paul and Kotliar}(2003)}]{pau03}
\bibinfo{author}{\bibfnamefont{I.}~\bibnamefont{Paul}} \bibnamefont{and}
  \bibinfo{author}{\bibfnamefont{G.}~\bibnamefont{Kotliar}},
  \bibinfo{journal}{Phys. Rev. B} \textbf{\bibinfo{volume}{67}},
  \bibinfo{pages}{115131} (\bibinfo{year}{2003}).

\bibitem[{\citenamefont{Small et~al.}(2003)\citenamefont{Small, Perez, and
  Kim}}]{sma03}
\bibinfo{author}{\bibfnamefont{J.~P.} \bibnamefont{Small}},
  \bibinfo{author}{\bibfnamefont{K.~M.} \bibnamefont{Perez}}, \bibnamefont{and}
  \bibinfo{author}{\bibfnamefont{P.}~\bibnamefont{Kim}},
  \bibinfo{journal}{Phys. Rev. Lett.} \textbf{\bibinfo{volume}{91}},
  \bibinfo{pages}{256801} (\bibinfo{year}{2003}).

\bibitem[{\citenamefont{Dresselhaus and Eklund}(2000)}]{dre00R}
\bibinfo{author}{\bibfnamefont{M.~S.} \bibnamefont{Dresselhaus}}
  \bibnamefont{and} \bibinfo{author}{\bibfnamefont{P.~C.}
  \bibnamefont{Eklund}}, \bibinfo{journal}{Adv. Phys.}
  \textbf{\bibinfo{volume}{49}}, \bibinfo{pages}{705} (\bibinfo{year}{2000}).

\bibitem[{\citenamefont{Graf et~al.}(1996)\citenamefont{Graf, Yip, Sauls, and
  Rainer}}]{graf96}
\bibinfo{author}{\bibfnamefont{M.~J.} \bibnamefont{Graf}},
  \bibinfo{author}{\bibfnamefont{S.-K.} \bibnamefont{Yip}},
  \bibinfo{author}{\bibfnamefont{J.~A.} \bibnamefont{Sauls}}, \bibnamefont{and}
  \bibinfo{author}{\bibfnamefont{D.}~\bibnamefont{Rainer}},
  \bibinfo{journal}{Phys. Rev. B} \textbf{\bibinfo{volume}{53}},
  \bibinfo{pages}{15147} (\bibinfo{year}{1996}).

\bibitem[{\citenamefont{{L\"ofwander} and {Fogelstr\"om}}(2004)}]{LF04}
\bibinfo{author}{\bibfnamefont{T.}~\bibnamefont{{L\"ofwander}}}
  \bibnamefont{and}
  \bibinfo{author}{\bibfnamefont{M.}~\bibnamefont{{Fogelstr\"om}}},
  \bibinfo{journal}{Phys. Rev. B} \textbf{\bibinfo{volume}{70}},
  \bibinfo{pages}{024515} (\bibinfo{year}{2004}).

\bibitem[{\citenamefont{{L\"ofwander} and {Fogelstr\"om}}(2005)}]{LF05}
\bibinfo{author}{\bibfnamefont{T.}~\bibnamefont{{L\"ofwander}}}
  \bibnamefont{and}
  \bibinfo{author}{\bibfnamefont{M.}~\bibnamefont{{Fogelstr\"om}}},
  \bibinfo{journal}{Phys. Rev. Lett.} \textbf{\bibinfo{volume}{95}},
  \bibinfo{pages}{107006} (\bibinfo{year}{2005}).

\end{thebibliography}

\end{document}